\documentclass[a4paper,8pt]{article}

\usepackage{a4wide}
\usepackage{authblk}
\usepackage{cite}
\usepackage[british]{babel}
\usepackage[utf8]{inputenc}
\usepackage{xcolor}
\usepackage{graphicx,subfigure}
\usepackage{epstopdf}
    \graphicspath{{./}{figures/}}
\usepackage[colorlinks=true,linkcolor=blue,citecolor=blue]{hyperref}
\usepackage{amsfonts,amsmath,amsthm}

\theoremstyle{remark}

\newcommand{\z}{\mathbf{z}}

\newcommand{\norm}[1]{\|#1\|}

\newcommand{\R}{\mathbb{R}}

\newcommand{\dd}{\mathrm{d}}
\newcommand{\disp}{\displaystyle}

\allowdisplaybreaks

\begin{document}

\title{Predictability of viral load {dynamics} in the early phases of SARS-CoV-2 through a model-based approach}

\author[1]{Andrea Bondesan}
\author[4]{Antonio Piralla}
\author[3]{Elena Ballante}
\author[4]{Antonino Maria Guglielmo Pitrolo}
\author[3]{Silvia Figini}
\author[4,5]{Fausto Baldanti}
\author[2]{Mattia Zanella \vspace*{3mm}}

\affil[1]{\normalsize Department of Mathematical, Physical and Computer Sciences, \protect\\ University of Parma, Parma, Italy \vspace*{2mm}}
\affil[2]{Department of Mathematics ``F. Casorati'', \protect\\ University of Pavia, Pavia, Italy \vspace*{2mm}} 
\affil[3]{Department of Political and Social Sciences, \protect\\ University of Pavia, Pavia, Italy \vspace*{2mm}} 
\affil[4]{Microbiology and Virology Department, \protect\\ Fondazione IRCCS Policlinico San Matteo, Pavia, Italy \vspace*{2mm}}
\affil[5]{Department of Clinical, Surgical, Diagnostic and Pediatric Sciences, \protect\\ University of Pavia, Pavia, Italy}
   
\date{}

\maketitle

\vspace*{-5mm}
\begin{abstract}
A pipeline to evaluate the evolution of viral dynamics based on a new model-driven approach has been developed in the present study. The proposed methods exploit real data and the multiscale structure of the infection dynamics to provide robust predictions of the epidemic dynamics. We focus on viral load kinetics whose dynamical features are typically available in the symptomatic stage of the infection. Hence, the epidemiological evolution is obtained by relying on a compartmental approach characterized by a varying infection rate to estimate early-stage viral load dynamics, of which few data are available. We test the proposed approach with real data of SARS-CoV-2 viral load kinetics collected from patients living in an Italian province. The considered database refers to early-phase infections, whose viral load kinetics have not been affected by the mass vaccination policies in Italy. 
\end{abstract}





\section{Introduction}

Infection dynamics are typically shaped by heterogeneous and interconnected factors, including exogenous individual-based aspects such as the individual response to threat, the collective compliance to non-pharmaceutical interventions, and variable endogenous characteristics of a pathogen \cite{BBSG,Brauer,BBT,BDM,Chen,Gatto_etal,Giordano,KDTHM,PME,Z_etal}. Indeed, the infectivity of individuals may vary strongly from the onset of the infection to its end. In a purely data-oriented approach the day of first contact with an infected individual is typically unknown or unreported and the infectivity levels may vary in time. Therefore, the trend of an epidemic naturally involves a plurality of timescales: one that is linked to the evolution of infected cases among the population and a second related with the change of infectivity related to in-host viral dynamics \cite{J_etal,LGT,BPT}.

In this direction, viral load (VL) dynamics represent a significant aspect influencing the disease progression and transmission. The evolution of VL constitutes a quantitative marker for assessing viral kinetics \cite{Frediani_etal,P_etal,V}. Nevertheless, the reconstruction of these trajectories is a challenging problem since data are incomplete and biased toward the peak viral load, which is typically related to the onset of symptoms in infected patients. We mention in this direction the empirical studies, see e.g. \cite{H_etal}. Hence, VL kinetics are strongly affected by uncertainties stemming from the data assimilation processes. Existing methods are not well-suited to consider possible mutations of the virus, resulting in a modification of the infectiousness curve associated with each individual illness. Therefore, a purely data-oriented approach should be complemented by a multiscale model-oriented approach to provide robust long-term predictions of epidemiological dynamics \cite{K_etal}. In the last few years, several mathematical approaches have been designed to assess the impact of infectivity dynamics at the population level \cite{DMLT,LT}. These works are based on kinetic-type equations and on simplified dynamics for which observable quantities are analytically computable. The developed methods are capable to link agent-based dynamics to available data, providing effective transition rates in macroscopic compartmental models, see also \cite{BBBEPT,BBDP,BTZ,DPTZ,FMZ,Z_BMB,Z_etal}.  

In this study, we focus on the definition of a new pipeline capable of extracting information starting from VL kinetics through a model-driven approach that couples the multiple timescales of infection dynamics. In detail, we exploit VL kinetics obtained in a series of SARS-CoV-2 infections that are not affected by exogenous influencing factors in the dynamics, like e.g. the implementation of mass vaccination policies. We interface data-oriented VL dynamics with a simple compartmental model involving age of infection \cite{DGMM,YZW} to describe the evolution of infected cases from SARS-CoV-2 in Italy. A new approach to evaluate the dynamics of the viral load has been developed and compared with original data obtained during the initial waves of the SARS-CoV-2 pandemic. 

Comprehending the factors influencing the in-host viral dynamics represents a fundamental tool to provide robust public health strategies. This pilot study could be implemented in further investigations involving other respiratory viruses, to better clarify the process of viral dynamics as a preparatory action for future pandemics.


\section{{Data and modeling assumptions}}\label{sect:2}

\subsection{Samples and SARS-CoV-2 kinetics}

Since February 2020, Fondazione IRCCS Policlinico San Matteo (Pavia, Italy) has been identified as reference regional center for the diagnosis of SARS-CoV-2 infections through the analysis of respiratory samples. In these analyses, RNA was extracted by using the MGISP-960 automated workstation and the MGI Easy Magnetic Beads Virus DNA/RNA Extraction Kit (MGI Technologies, Shenzhen, China). SARS-CoV-2 RNA was detected using the SARS-CoV-2 variants ELITe MGB kit (ELITechGroup, Puteaux, France; cat. no. RTS170ING) targeting ORF8 and RdRp gene. Reactions were carried out on the CFX96 Touch Real-time PCR detection system (BioRad, Mississauga, ON, Canada).

Individuals that resulted positive from SARS-CoV-2 RNA may leave isolation only when two consecutive negative respiratory samples occurred. Thus, a series of follow-up samples were analyzed to evaluate the structure of SARS-CoV-2 viral loads and used to monitor the evolution of SARS-CoV-2 VL kinetics. {For this reason, the VL kinetics of each considered patient are described by at least two samples, one accounting for the time of testing and the other for the measured viral load. In more detail,} the experimental data for each patient consist of two $n$-tuples, $n \in \mathbb{N}$, where $n$ is the number of successive molecular tests taken by the individual during the monitoring period: the first $n$-tuple contains the times (in days) when respiratory samples were collected and analyzed (where 0 always represents the day of the first test) and the second contains the kinetics $\textrm{Ct}$ detected by each of these tests (where $40\ \textrm{Ct}$ means complete negativity to the virus and thus recovery). So, for example, a patient that took three successive molecular tests, the first being strongly positive ($\sim 20\  \textrm{Ct}$), the second barely positive ($\sim 36\  \textrm{Ct}$) after 9 days and finally the third negative after 7 more days, would be characterized by the two triples $(0,9,16)$ for the times and $(20,36,40)$ for the kinetics. The values of the corresponding viral loads in terms of RNA $\textrm{copies}/\textrm{ml}$ of respiratory samples may then be deduced from the kinetics {of cycle threshold ($\textrm{Ct}$)} using the conversion formula {for the quantification of copy number in clinical samples proposed in \cite{PD,Pir2}, which is based on the following proportion  }
\begin{equation}
\label{eq:conversion}
\disp 1\ \frac{\textrm{copies}}{\textrm{ml}} = 16.6 \times 10^{- 0.2814\ \times\ 1 \textrm{Ct}\ +\ 11.232},
\end{equation}
where $16.6$ is the dilution factor needed to report VL as RNA $\textrm{copies}/\textrm{ml}$ of respiratory samples. Laboratory data were anonymized and retrospectively analyzed. The study was conducted in accordance with the Declaration of Helsinki and no clinical information were available except for the age and gender. 

\subsection{Statistical analysis}

Numerical variables are described as mean $\pm$ standard deviation, categorical variables as count and percentages. The comparison of the two time periods {(November--December 2020 and January--May 2021)} considered is performed in terms of $t$-test for quantitative variables and Fisher exact test for quantitative variables. Considering the longitudinal structure of the data, they are presented as daily averaged values. The comparison between the two periods and the investigation of the influence of age and gender on the kinetics are performed in terms of a linear mixed-effects model. The temporal evolution of VL kinetics is obtained as a Gamma distribution using the Matlab built-in \texttt{fit} function, with the LAR method to ensure that the extreme values have a lesser influence on the result. The significance threshold is set to $p < 0.05$.

{
\subsection{Data description}
The study analyzed a total of 233 VL kinetics, corresponding to approximately 700 respiratory samples, collected from patients across two distinct periods. The first period goes from November to December 2020, before circulation of any variants of concern (VOCs), and included 71 patients. The second period, going from January to May 2021 during the circulation of VOCs, involved 162 patients. The subjects considered had a mean age of $42.3 \pm 14.7$ years, with $55 \%$ of females (120) and $45 \%$ of males (113). The demographic distribution, in terms of age and gender, does not statistically differ between the two periods. Using mixed-effects models we found that age and gender did not significantly influence VL kinetics. Time to negativization has a significant effect, showing a decreasing pattern of VL kinetics ($\textrm{estimate} = -482193.7$, $\textrm{p-value} < 0.001$). Additionally, a significant difference in VL kinetics was observed between the two periods, with lower values recorded in 2021 compared to 2020 ($\textrm{estimate} = -2122031.9$, $\textrm{p-value} = 0.015$).
}

{
\subsection{Viral load kinetics}
In this section we propose a reconstruction based on raw data of VL kinetics, to understand the main features of their evolution. Based on existing works in the medical literature about the shape of the SARS-CoV-2 infectiousness \cite{CevKupKinPei, PME} we expect the viral load to be characterized by a rapid growth in the few days after the first contact with an infective individual, until the infection reaches its maximum, followed by a slower decay up to the complete negativization of the subject after several days. {We recall that a patient is considered negative if $\textrm{Ct} \ge40$ is detected, based on the conversion formula \eqref{eq:conversion}. In order to compare VL kinetics from different patients, we {determine} the first day of negativization for each subject and we save all their preceding, patient-specific, VL measurements.} {Since there is no a priori information on the day when the peak of infectivity is reached, we align the different VL kinetics of all subjects based on the day of negativization, being this value determined within the whole dataset. In particular, the considered alignment does not depend on the observed peak of individual viral loads, which in turn is often the first measurement. Therefore, an alignment by negativization makes possible to observe an aggregate trend in the dataset. } 
In Figure \ref{fig:VL raw data} {we present the raw data of VL kinetics for the whole cohort of patients, aligned by the day of first negativization}, over the two distinct time periods during which their infectiousness was monitored.
\begin{figure}
\centering  
\includegraphics[scale=0.4]{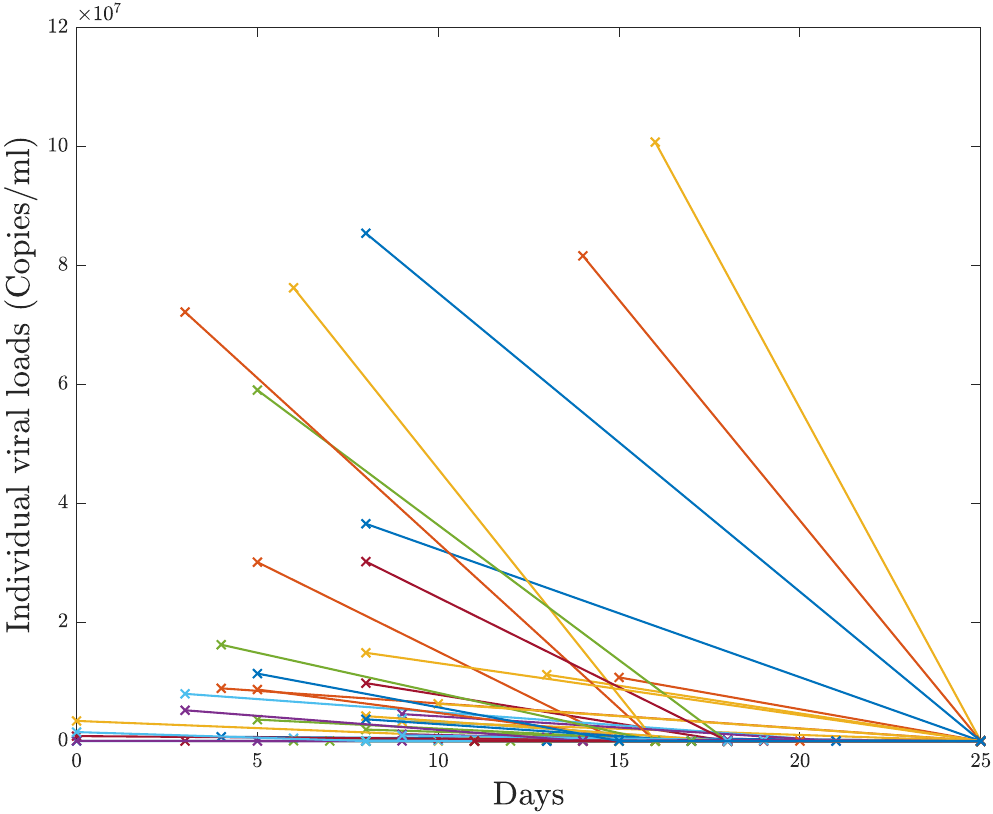} \hspace*{5mm}
\includegraphics[scale=0.4]{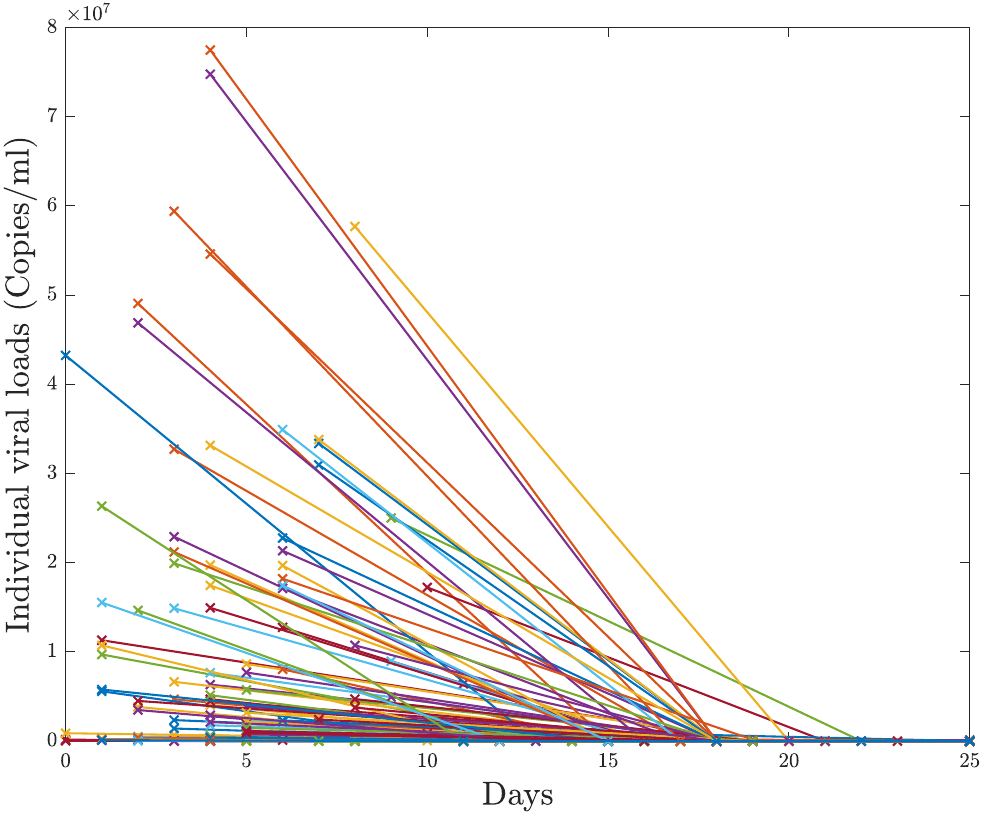}
 \caption{Evolution of individuals' viral load kinetics over time with an ordering based on the first negative test taken by the infective patients, where a patient is considered negative if $\textrm{Ct} \ge 40$ is detected, based on the conversion formula \eqref{eq:conversion}. {We use a simple linear interpolation to connect values that are relative to the same subject. } We distinguish between the two observation periods of November--December 2020, accounting for 71 patients (left), and January--May 2021, accounting for 162 patients (right).}
\label{fig:VL raw data}
\end{figure}
{Based on this alignment process, we consider the daily average of the viral load with respect to the number of patients that got tested on the corresponding day.} We then infer the evolution of the viral load as the best fit of these averages, weighted by the daily patients' count. This means that the more patients got tested on one specific day, the more importance the corresponding average VL kinetics will have in the fitting. The outcomes of this analysis are shown in Figure \ref{fig:Viral load fit}. The infection persists for at most 25 days and its peak is reached 3 to 4 days after the first contact with an infected individual (4 days for the viral loads considered from the period November--December 2020 and 3 days for those belonging to the window January--May 2021). The red curves represent the best weighted fits for the daily averaged viral loads and have the shape of a Gamma distribution
\begin{equation} \label{eq:VL_shape}
\hat \beta(\tau) = \alpha_0 (\tau + \alpha_1)^{\alpha_2}\exp\left\{-\alpha_3(\tau + \alpha_1)\right\}, 
\end{equation}
where $\alpha_0 = 10^7$ is a scale factor used to capture the magnitude of individuals' viral load and the coefficients $\alpha_1 \in \R$, $\alpha_2$, $\alpha_3 \geq 0$ are reported in Table \ref{tab:Optimal coeffs}. In particular, $\alpha_1$ represents the time of first contact (measured in days), while $\alpha_2$ and $\alpha_3$ characterize the shape of the viral load, hence the probability of viral transmission between individuals.
\begin{figure}[h!]
\centering  
\includegraphics[scale=0.4]{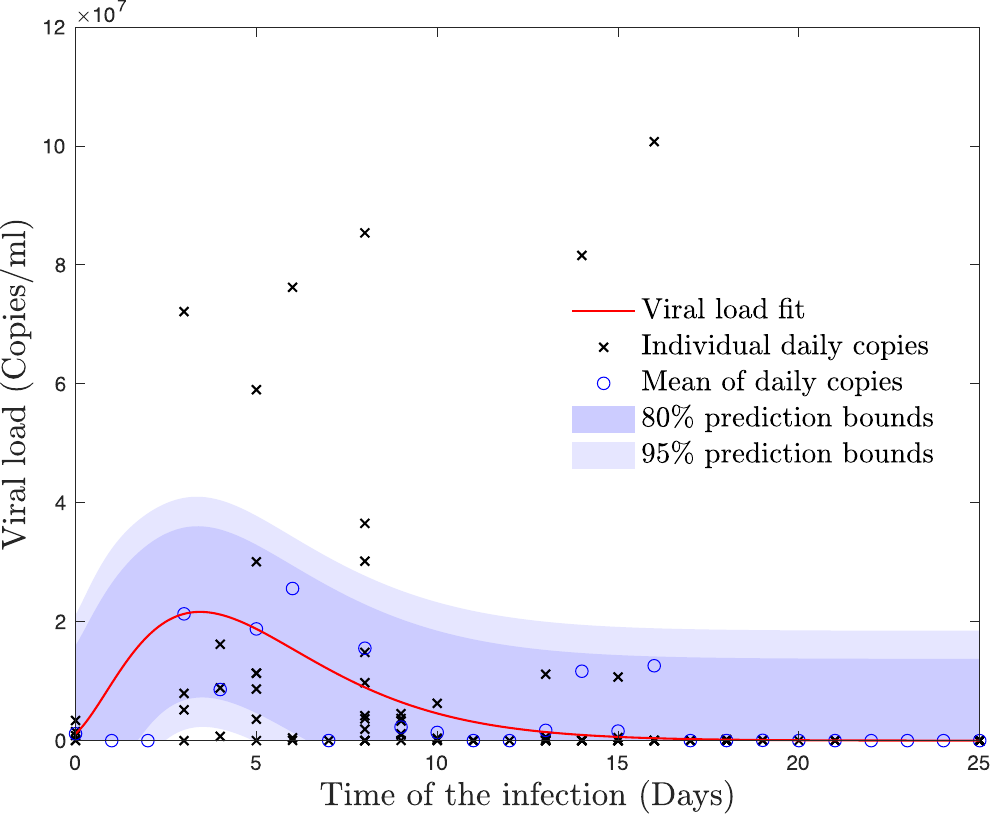} \hspace*{5mm}
\includegraphics[scale=0.4]{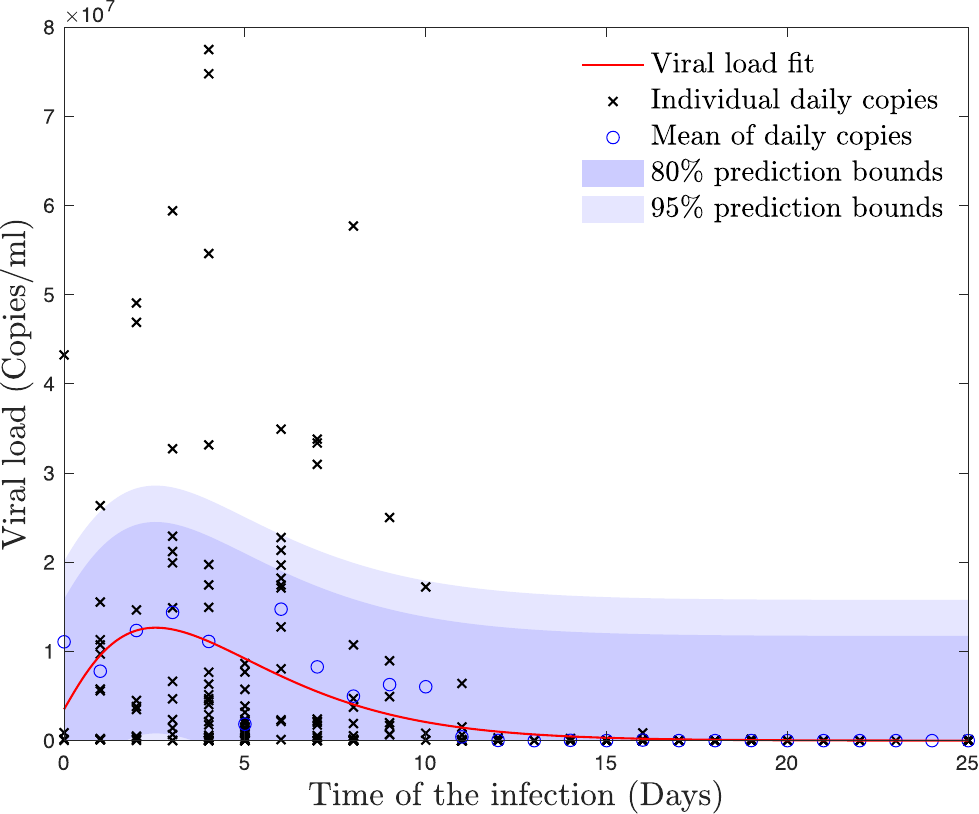}
\caption{Best fit of the patients' viral loads depending on the time of the infection, for the two monitoring periods of November--December 2020 (left) and January--May 2021 (right). The black crosses denote all the individuals' infection cycles ordered based on their first negativization. The blue dots represent the averages of the daily viral loads. The red curves are the best fit for these averages, weighted by the number of patients counted on each day. The shaded region gives the $80 \%$ and $95 \%$ prediction bounds for the whole dataset.}
\label{fig:Viral load fit}
\end{figure}
\begin{table}
\centering
\begin{tabular}{|c|c|c|}
\hline
\multicolumn{3}{|c|}{November--December 2020's sample} \\
\hline
                  & 	Estimated value 	&	 $95\%$ CI  		\\
\hline
$\alpha_1$ &	0.4189 d			&	[-0.4865, 1.3245]	\\
$\alpha_2$ &	2.2015			&	[1.6124, 2.7906]	\\
$\alpha_3$ &	0.5701			&	[0.3928, 0.7475]	\\
\hline
\end{tabular} \hspace*{5mm}
\begin{tabular}{|c|c|c|}
\hline
\multicolumn{3}{|c|}{January--May 2021's sample} \\
\hline
                  & 	Estimated value 	& 	$95\%$ CI 		\\
\hline
$\alpha_1$ &	0.6262 d			&	[0.1588, 1.0936]	\\
$\alpha_2$ &	1.5687			&	[1.0462, 2.0912]	\\
$\alpha_3$ &	0.4964			&	[0.3348, 0.6580]	\\
\hline
\end{tabular}
\caption{Optimised coefficients for viral load kinetics, for the two observation periods of November--December 2020 (left) and January--May 2021 (right). The parameter $\alpha_1$ denotes the time of first contact (measured in days) while the shape parameters $\alpha_2$ and $\alpha_3$ characterize the probability of transmitting the infection.}
\label{tab:Optimal coeffs}
\end{table}
}

\subsection{Epidemic model with uncertain quantities}

Compartmental models are classically defined to describe mathematically the spread of epidemics in a population \cite{DH}. The main idea of such models is to divide the population into different subgroups, each identified by a specific epidemiologically relevant status. Suitable transition rates are introduced to describe the switches between compartments. Generalizations of the mentioned compartmental models include additional factors characterizing the transmission dynamics such as external influence, age structure, variable contact dynamics and mobility of agents, see e.g. \cite{APZ,BBPSV,CS,DT}. {From a model perspective, the choice of the compartmentalization is crucial to provide reliable predictions and, therefore, to effectively understand the evolution of the infection. In the following, we will adopt a SIR-type model with age of infection, where the compartment propagating the infection is the one identified by the population of infected agents. In particular, assuming the infectivity to be modulated by an additional variable identifying the age of infection, we can keep track of the disease's variability inside each individual, as it evolves in parallel with the main timescale of the epidemic. In other words, we assume that the viral load heterogeneity is capable to characterize the population propagating the infection \cite{J_etal}. Experimental evidence suggests that in the growth phase of VL dynamics no symptoms are typically detected, see \cite{M_etal}.   } This allows to introduce a local incidence rate function of the age of infection, modeling the average viral load of the infected individuals and whose shape can be inferred by the information coming from real data on molecular tests. To take into account the structural uncertainties stemming from data assimilation processes of VL kinetics, we extend the model approach to consider the presence of uncertain quantities as a structural feature of the epidemic dynamics. To date, few results are available regarding the development of uncertainty quantification (UQ) methods in epidemic systems, see \cite{APZ,CCVH,Chowell,R}. The development of UQ methods are based on an increased dimensionality of the problem, that is made dependent on a random variable $\mathbf z \in \mathbb R^{d_z}$ of which the distribution $p(\z)$ is known  \cite{JP,Par,Xiu}. The extrapolation of statistics is classically obtained by looking at quantities of interest, with respect to $\mathbf{z}$, representing the expected solution of the problem. 

{Since the first day of infectivity of an infectious individual is unknown, we consider a compartmental modeling approach with uncertainty.  } We consider two temporal scales, one for the time $t \geq 0$ of evolution for the epidemic and another for the time $\tau \geq 0$ of evolution for the infection. At any given time $t \geq 0$, we then {divide} the population into susceptible individuals $S = S(t,\z)$, recovered individuals $R = R(t,\z)$ and infected individuals $i(t,\tau,\z)$, with variable infection over time $\tau \geq 0$. We then consider the triple $(S(t,\z),i(t,\tau,\z),R(t,\z))$ of unknowns, solution to 
\begin{equation} \label{SIR-AOI}
\left\{ \begin{array}{l}
\disp \frac{\dd S(t,\z)}{\dd t} = - \bar{\beta} S(t,\z) \int_{\R_+} \beta(\tau,\z) i(t,\tau,\z) \dd \tau, \\[5mm]
\disp \partial_t i(t,\tau,\z) + \partial_\tau i(t,\tau,\z) = - \bar{\gamma} \gamma(\tau,\z) i(t,\tau,\z), \\[5mm]
\disp \frac{\dd R(t,\z)}{\dd t} = \bar{\gamma} \int_{\R_+} \gamma(\tau,\z) i(t,\tau,\z) \dd \tau,
\end{array} \right.
\end{equation}
where $\beta(\tau,\z) \geq 0$ and $\gamma(\tau,\z) \geq 0$ represent respectively the uncertain infection {rate} and the uncertain recovery {rate} of the infected individuals, depending on the age of infection $\tau \geq 0$, while the constants $\bar{\beta} > 0$ and $\bar{\gamma} > 0$ enclose respectively the baseline contact rate and the {baseline recovery rate}. {In \eqref{SIR-AOI} we indicated with $\partial_t$ the partial derivative with respect to the epidemic timescale $t \geq 0$ and with $\partial_\tau$ the partial derivative computed with respect to the infection timescale $\tau \geq 0$, see \cite{DGMM}. In particular, we stress once more that} since the observable number of infected individuals also varies based on the age of infection $\tau\ge0$, the quantity $i(t,\tau,\z) \geq 0$ depends on both timescales (epidemic and infection course) and is related to the total current number of infected individuals $I = I(t,\z)$ through the relation 
\[  \displaystyle I(t,\z) = \int_{\R_+} i(t,\tau,\z) \dd \tau.  \]
Therefore, this model couples the timescale of the epidemic with the one characterizing the evolution of the virus in terms of its average infectiousness. 
For each $\z$, the system is finally completed with suitable initial and boundary conditions for the variables $S(t,\z)$, $i(t,\tau,\z)$ and $R(t,\z)$. A natural choice is
\begin{gather}
S(0) = S_0, \quad i(0, \tau,\z) = i_0(\tau,\z), \quad R(0) = R_0,  \label{IC} \\[5mm]
i(t, 0,\z) = \bar{\beta} S(t,\z) \int_{\R_+} \beta(\tau,\z) i(t,\tau,\z) \dd \tau, \label{BC}
\end{gather}
where $S_0$, $i_0(\tau,\z)$, $R_0 \in \R_+$ are such that  
\[  \displaystyle S_0 + \int_{\mathbb R^{d_z}}\int_{\R_+} i_0(\tau,\z) \dd \tau \dd p(\z) + R_0 = 1,  \]
and $p(\z)$ is the distribution of the random variable $\z$. 

{The introduced simplified infection dynamics may be further modified to take into account a more complex compartmentalization of the society, the influence of external interventions and infection delays among others, see \cite{BDM,FIV,Giordano} and the references therein. In the following, we will always assume a constant $\gamma(\tau,\mathbf z) \equiv 1$, so that the recovery rate is simply given by $\bar{\gamma}$. We point the interested reader to \cite{Z_etal} for a more precise estimation of such parameters.  }


\section{{Inference of viral load dynamics} }\label{sect:3}
In this section we define a pipeline to effectively estimate VL {dynamics} $\beta(\tau,\z)$ by using available epidemiological data and introducing uncertainty in the day of first contact, expressed by the coefficient $\alpha_1$ in \eqref{eq:VL_shape}. To this end, we consider an uncertain parameter $\alpha_1(\z) \in \R$ to make the local incidence rate random. Moreover, we take into account data from the COVID-19 pandemic in Italy, at both local and national level, by first considering the evolution of infected individuals in the province of Pavia and by then extending our analysis to the whole territory of Italy, where the central government implemented several non-pharmaceutical interventions (NPIs) during the pandemic. In Table \ref{tab:Main NPIs} we report the timeline of the main NPIs taken before the mass vaccination campaign started in mid 2021. It is however worth mentioning that local additional measures, like isolation of small portions of a territory, had often been implemented as well.

Before advancing to the estimation of VL dynamics, it is crucial to address some key points. We emphasize the inherent complexity in calibrating epidemiological models, which is particularly demanding from a numerical perspective. Indeed, the process of aligning these models with real-world scenarios requires a careful approach to parameter estimation. Furthermore, available data on VL kinetics tend to hide the initial evolution of their dynamics since agents are typically tested after the manifestation of the symptoms. This bias is primarily attributed to limited testing capacity, resulting in a dataset that often reflects a conservative estimate. Consequently, any attempts to determine the VL dynamics should be approached with a mindful consideration of these intrinsic limitations, coupled with a commitment to refining the model's accuracy through continuous reassessment and adaptation.

\begin{table}[h]
\begin{center}
\begin{tabular}{|c|c|c|}
\hline
Implemented NPI			&	Date			&	Average recovery {time} \\
\hline \hline
First lockdown				& March 2020	&					   \\
Curfew measures			& October 2020	&	14 d 		   		   \\
Strengthening of measures	& March 2021	&					   \\
 \hline
\end{tabular}
\end{center}
\caption{Timeline of the main NPIs employed in Italy before the mass vaccination campaign started in mid 2021. The average recovery {time} from the infection was $14$ days.}
\label{tab:Main NPIs}
\end{table}

We shall focus on three different periods of time. We start by analyzing the first COVID-19 epidemic wave in Italy, which began in February--March 2020. We then proceed with the study of the second wave of October 2020 and conclude with the third major wave of February--March 2021. {We stress that the data on VL kinetics are limited to the sole province of Pavia and have been collected during the two periods of November--December 2020 (characterized by the absence of VOCs) and January--May 2021 (involving the circulation of VOCs).}  Moreover, we point out that the epidemiological data at our disposal for the province of Pavia were only acquired up to January 18 2021 and we will therefore narrow down our analysis of the third epidemic wave solely to the case of Italy, for which we shall instead take advantage of the open access data collected by the Johns Hopkins University (\href{https://github.com/CSSEGISandData/COVID-19}{github.com/CSSEGISandData/COVID-19}), reporting in detail the daily number of infected, recovered and deceased individuals in Italy up to August 4 2021. Let $\mathcal{C}_z = \{\z_1,\dots,\z_M\}$ be a sample of the random variable $\z$ having probability density function $p(\z)$. We adopt a multi-level approach to estimate the parameters $(\alpha_i)_{i=1,2,3}$ characterizing $\beta(\tau,\z_j) \geq 0$, which consists of two steps.

\smallskip
\noindent \textbf{Step 1 -- Determination of the epidemiological parameters.} For starting, we initially assess the (deterministic) value of $\alpha_1$ dictated by the epidemic during the two weeks preceding any of the three main NPIs presented in Table \ref{tab:Main NPIs} and we estimate at the same time the relevant epidemiological parameter $\bar{\beta} > 0$ which defines the baseline contact rate, assuming no restrictions on social contacts were in place. Notice that we have to discard the variable $\z$ at this stage, because we first need to ensure that the day of first contact $\alpha_1$ is compatible with the epidemiological data, in order to quantify the shift between the infection and the epidemic timescales. For this, we look for the optimal $\alpha_1$ and $\bar{\beta}$ minimizing the relative weighted $L^2$ norm between the current number of infected and recovered individuals (in Pavia or in Italy, depending on the local or national setting under consideration), which we denote $\hat{I}(t)$ and $\hat{R}(t)$, and the theoretical evolution of these quantities, $I(t) = \int_{\R_+} i(\tau) \dd \tau$ and $R(t)$, given by the model \eqref{SIR-AOI} in absence of uncertainty. We notice in particular that, since \eqref{SIR-AOI} only involves the recovered individuals inside the population, we take $\hat{R}(t)$ to be the sum of the reported numbers of recovered and deceased individuals. We thus solve the minimization problem
\begin{equation} \label{Minimization1}
\min_{\alpha_1, \bar{\beta}} \ (1-\eta) \frac{\norm{ I(t) - \hat{I}(t) }_{L^2([t_0,t_f])}}{\norm{ \hat{I}(t) }_{L^2([t_0,t_f])}} + \eta \frac{\norm{ I(t) + R(t) - \hat{I}(t) - \hat{R}(t) }_{L^2([t_0,t_f])}}{\norm{ \hat{I}(t) + \hat{R}(t) }_{L^2([t_0,t_f])}},
\end{equation}
over the time horizon $[t_0, t_f]$ representing any of the three periods that can be identified with the initial spreading phase of the three epidemic waves under study: February 24--March 9 2020, October 7--October 22 2020 and February 24--March 10 2021. In particular, we choose here the infection rate $\beta(\tau)$ to be the Gamma function \eqref{eq:VL_shape} obtained in the previous section from the viral loads measurements, rescaled by the factor $\alpha_0 = 10^7$. We use the fitting from November--December 2020 as initial guess for the optimization \eqref{Minimization1} during the first two epidemic waves, while we opt for the January--May 2021's fitting to study the third wave. Notice that at this stage the parameters $\alpha_2$ and $\alpha_3$ are fixed from these two fits, since we only need to ensure a proper initial calibration of the time shift $\alpha_1$ to correctly align our model with the epidemiological data. Finally, we always consider a constant recovery rate $\bar \gamma = 1/14$ to lower the complexity of the minimization. This value is consistent with the existing literature on the COVID-19 infection \cite{Chen,Lavezzo,Z_etal}, for which the time to viral clearance has been reported to span approximately between 10 and 21 days on average over the different variants. In Table \ref{tab:Epidemiological params} we collect all the parameters obtained from this initial analysis, for each of the three waves.

\begin{table}[h]
\centering
\begin{tabular}{|c|c|c|}
\hline
\multicolumn{3}{|c|}{Province of Pavia} \\
\hline
                   	  & First wave   				& Second wave  			\\
\hline
$\alpha_1$ 	  &	1.8110 d				&	-0.6706 d				\\
$\alpha_2$ 	  &	2.2015				&	2.2015				\\
$\alpha_3$ 	  &	0.5701				&	0.5701				\\
$\bar{\beta}$ 	  &	0.1486\ $\textrm{d}^{-1}$	&	0.1603\ $\textrm{d}^{-1}$  \\
$\bar{\gamma}$ &	1/14\ $\textrm{d}^{-1}$	&	1/14\ $\textrm{d}^{-1}$	\\
\hline
\end{tabular} \hspace*{5mm}
\begin{tabular}{|c|c|c|c|}
\hline
\multicolumn{4}{|c|}{Italy} \\
\hline
                  	 & First wave 				& Second wave 			& Third wave  				\\
\hline
$\alpha_1$ 	 &	6.6849 d				&	-0.6032 d  			&	 -0.8251 d				\\
$\alpha_2$ 	 &	2.2015				&	2.2015  				&	 1.5687				\\
$\alpha_3$ 	 &	0.5701				&	0.5701  				&	 0.4964				\\
$\bar{\beta}$ 	 &	0.3992\ $\textrm{d}^{-1}$	&	0.1225\ $\textrm{d}^{-1}$  	&	 0.1323\ $\textrm{d}^{-1}$	\\
$\bar{\gamma}$ &	1/14\ $\textrm{d}^{-1}$	&	1/14\ $\textrm{d}^{-1}$      &	 1/14\ $\textrm{d}^{-1}$	\\
\hline
\end{tabular}
\caption{Epidemiological parameters and coefficients of the infectiousness function for the three periods of time under consideration, both at a local (province of Pavia) and a national level (Italy).}
\label{tab:Epidemiological params}
\end{table}

\smallskip
\noindent \textbf{Step 2 -- Inclusion of the uncertainty.} We now go further in the analysis by including the presence of uncertainty and by looking at the full model \eqref{SIR-AOI}. The aim is to perform an uncertainty quantification on the shape of $\beta(\tau, \z_j)$ with respect to the randomness $(\z_j)_{j=1}^M$ that may appear when estimating $\alpha_1(\z_j)$, the day of first contact. For this, we perturb the optimal value $\alpha_1$ computed in the previous step by a centered uniform random distribution $\mathcal{U}([-2,2])$, i.e. $\alpha_1(\z_j) = \alpha_1 + \z_j$ with $\z_j \sim \mathcal{U}([-2,2])$, and we assess the impact of this addition on the parameters $\alpha_2(\z_j)$ and $\alpha_3(\z_j)$ that define the shape of $\beta(\tau,\z_j)$. The (randomized) optimal shape is thus determined by fitting the model \eqref{SIR-AOI} to reproduce the data on infected and recovered cases, through a minimization problem similar to \eqref{Minimization1}, but with parameters $\alpha_1(\z_j)$ and $\alpha_2(\z_j)$ instead of $\alpha_1$ and $\bar{\beta}$, while of course the latter and $\bar{\gamma}$ are those obtained from the previous step. This procedure is repeated for the whole sample $\mathcal{C}_z$ and we can finally infer the optimal shape of the infectiousness function with UQ to be the expected value $\mathbb{E}_z(\beta(\tau,\z))$ with respect to $\z$, based on the probability density $p(\z) = \mathcal{U}([-2,2])(\z)$.

\begin{figure}[h!]
\centering
\includegraphics[scale=0.4]{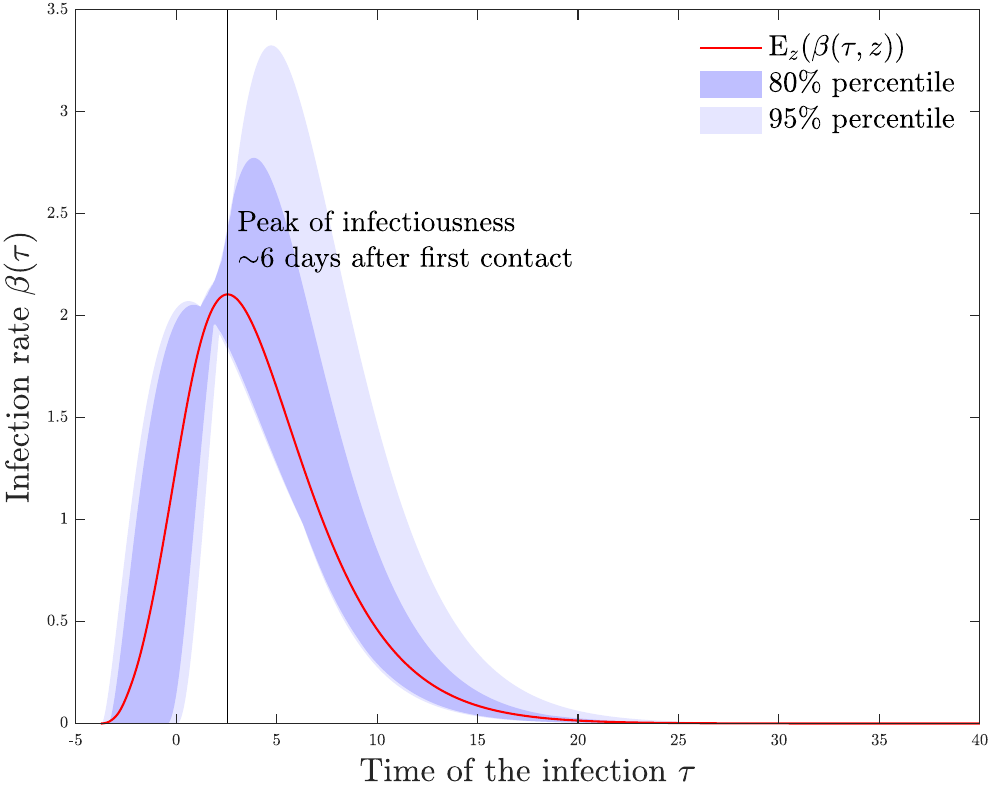} \hspace*{5mm}
\includegraphics[scale=0.4]{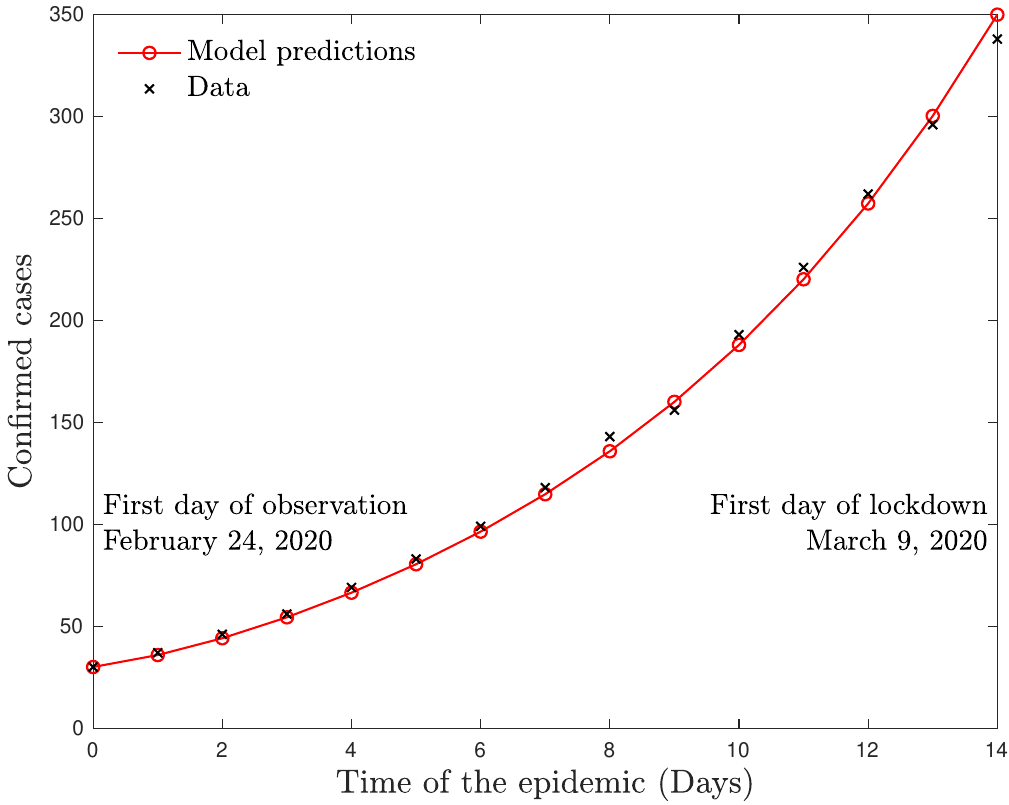}
\includegraphics[scale=0.4]{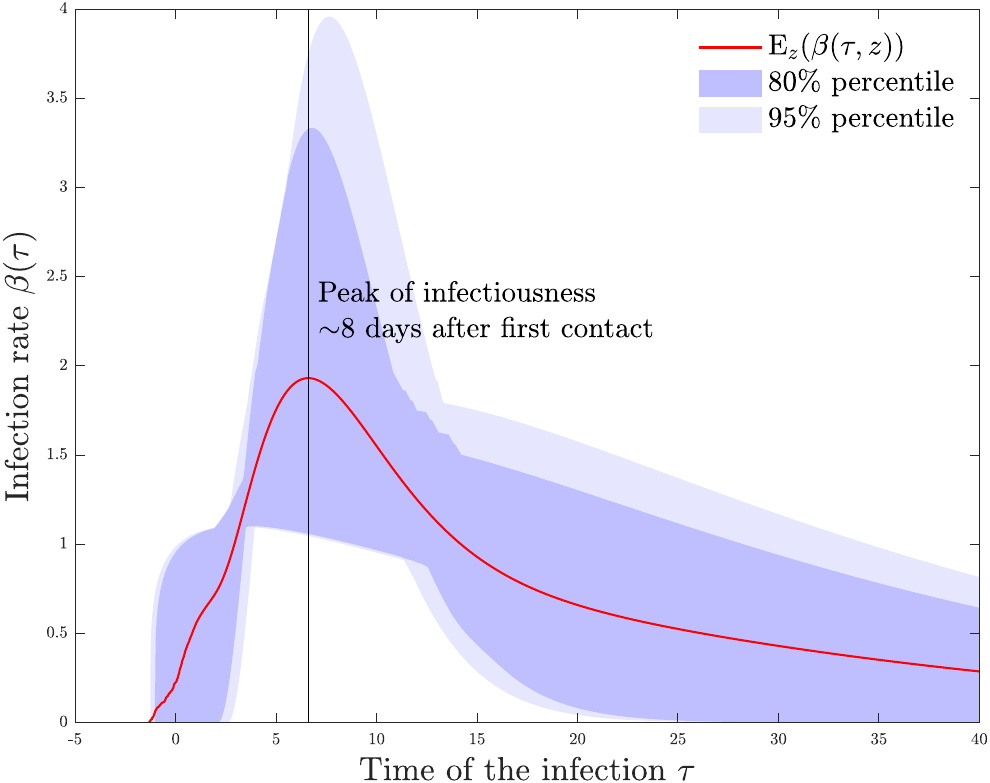} \hspace*{5mm}
\includegraphics[scale=0.4]{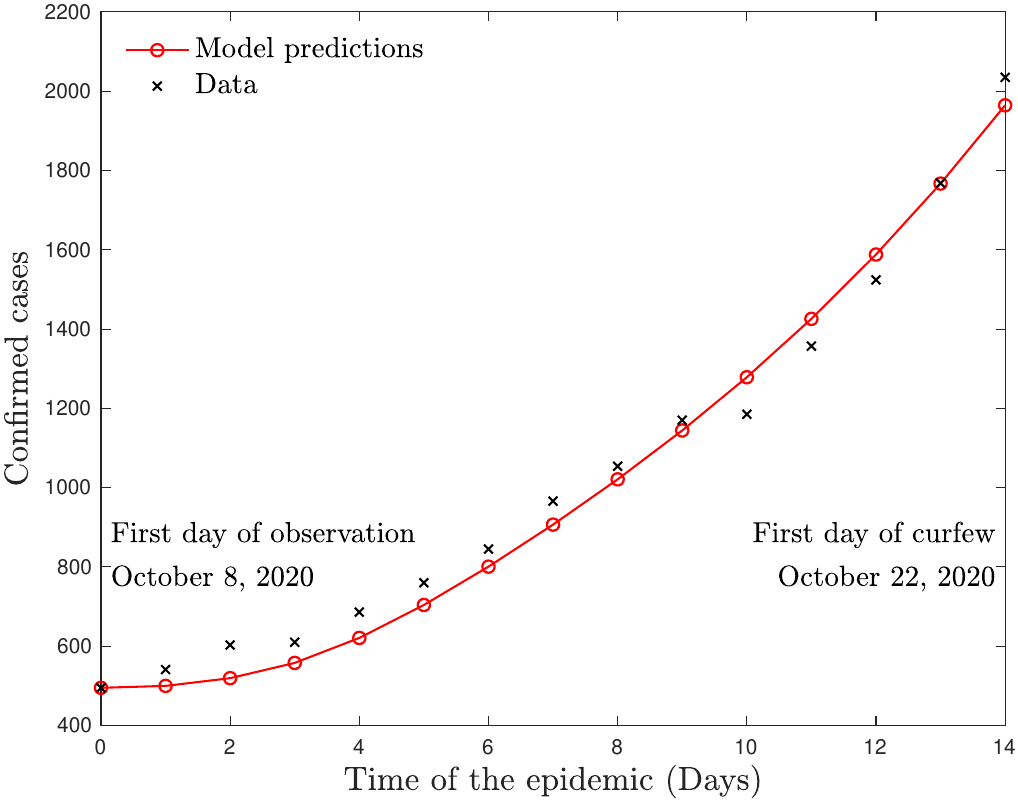}
 \caption{Data- and model-driven inference of the optimal shape of the infection rate in the province of Pavia (Italy) over two different epidemic waves: the one occurred at the beginning of the pandemic in 2020, with a focus on the two weeks (February 24--March 9) preceding the first lockdown (top), and the other occurred in the fall of the same year, with a focus on the two weeks (October 7--22) preceding the imposition of a national curfew (bottom). The figures on the left provide the optimal shapes $\mathbb{E}_z(\beta(\tau,\z))$ of the infectiousness function (red curves) determined for these two periods, starting from the data collected by Fondazione IRCCS Policlinico San Matteo to infer an initial guess of the infection rate through the function \eqref{eq:VL_shape} rescaled by the factor $\alpha_0 = 10^7$. The shaded areas show how the infectiousness function is altered when a uniform noise $\mathcal{U}([-2,2])$ modifies the day of first contact. The figures on the right present the evolution of infected individuals from SARS-CoV-2 in the province of Pavia over the considered windows of time. We compare computed (in red) and reported (in black) number of cases.}
 \label{fig:VL Pavia}
\end{figure}

\begin{figure}[h!]
\centering
\includegraphics[scale=0.4]{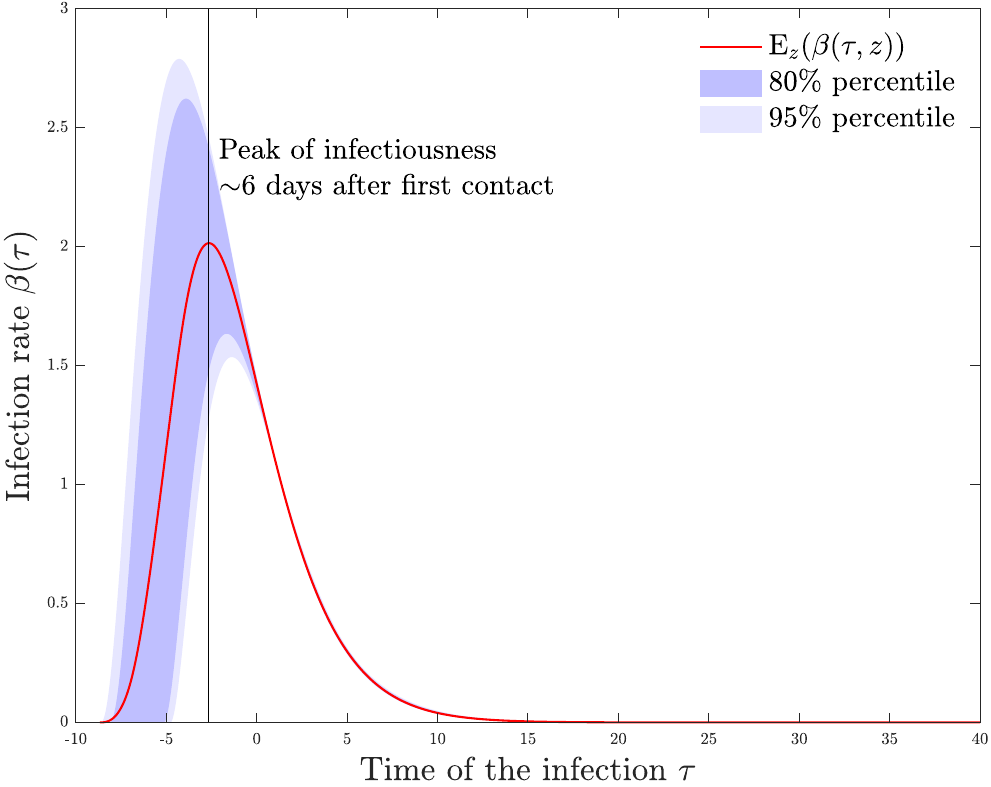} \hspace*{5mm} \includegraphics[scale=0.4]{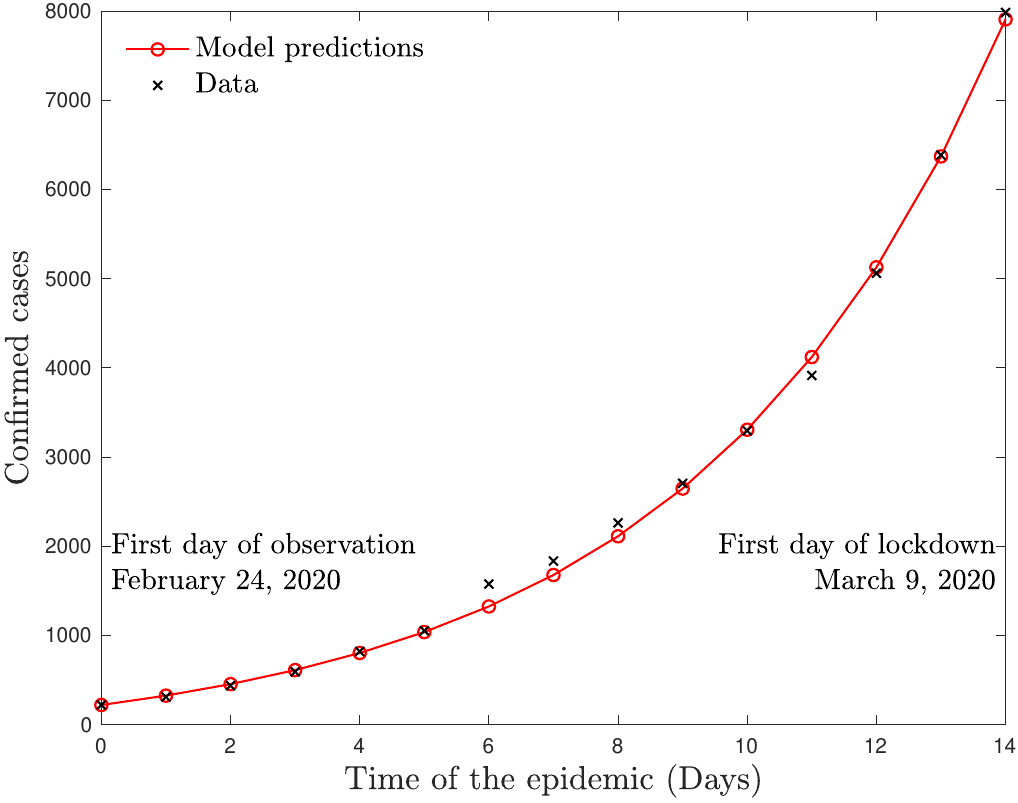}
\includegraphics[scale=0.4]{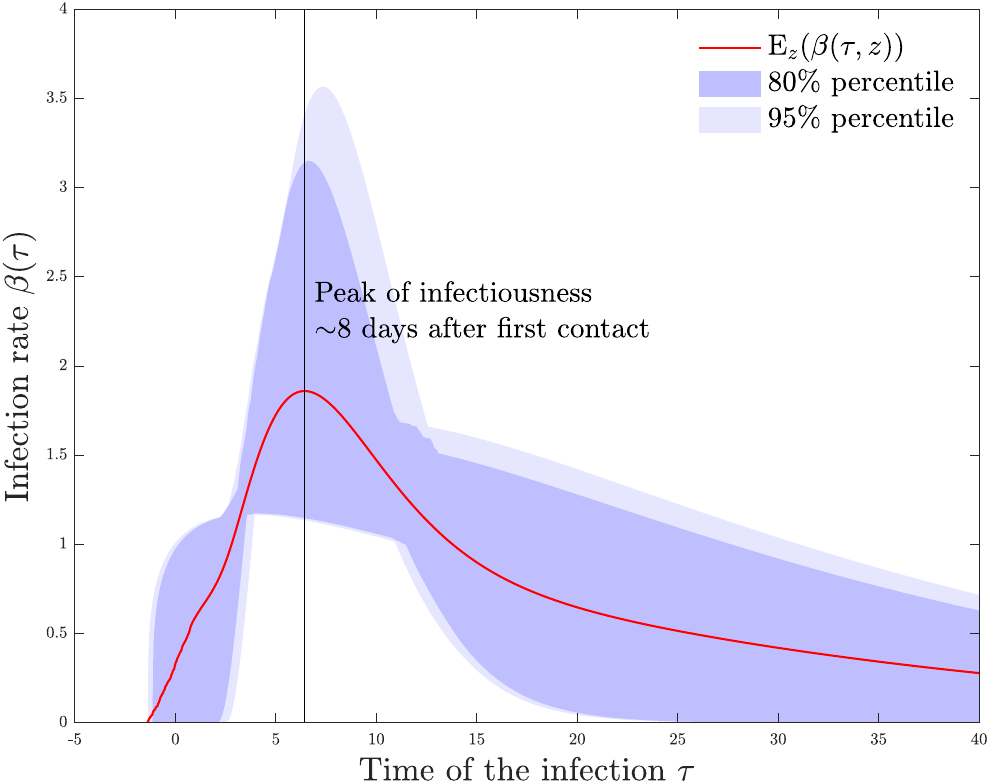} \hspace*{5mm} \includegraphics[scale=0.4]{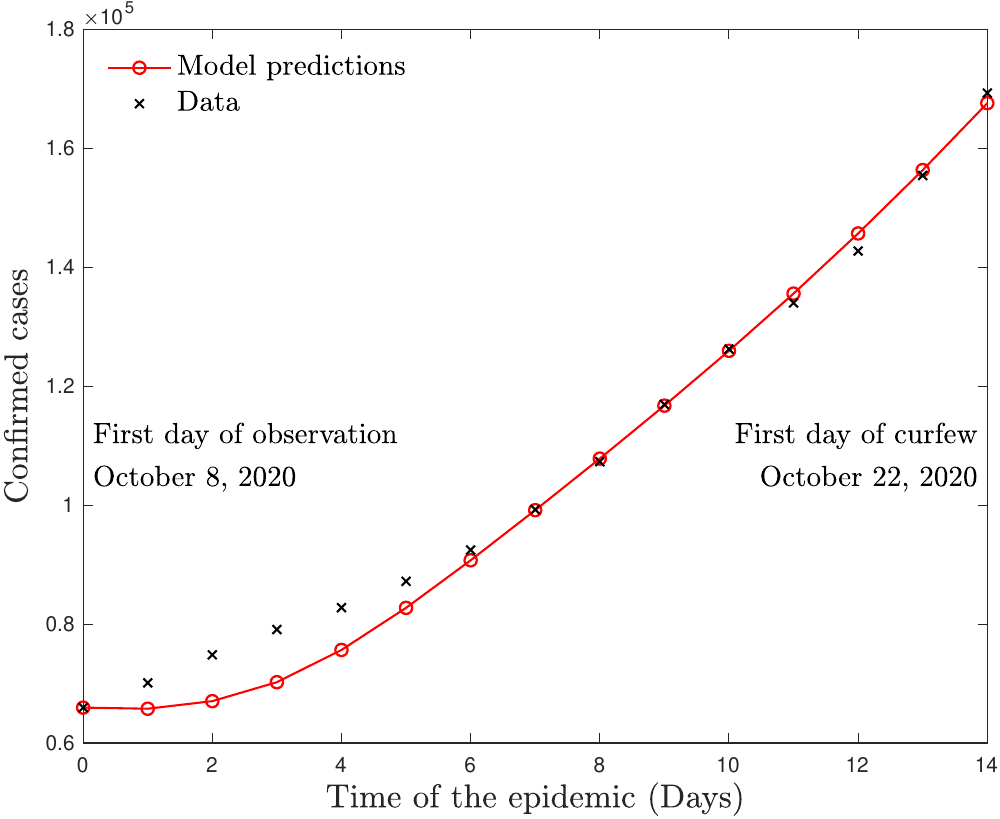}
\includegraphics[scale=0.4]{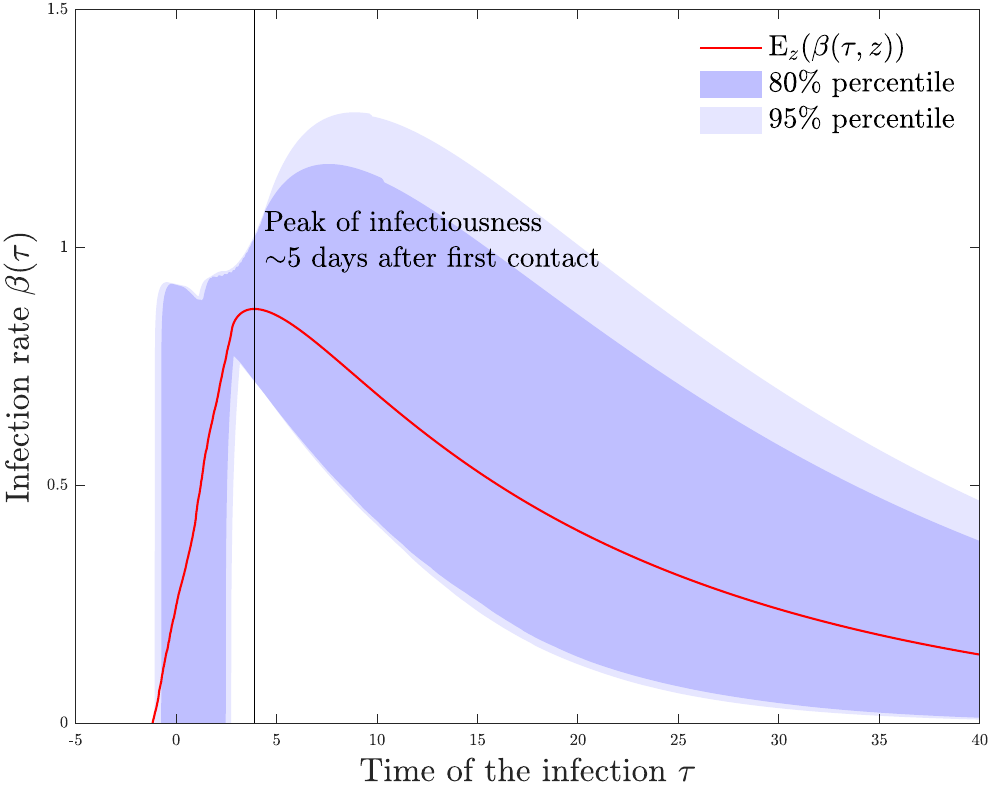} \hspace*{5mm} \includegraphics[scale=0.4]{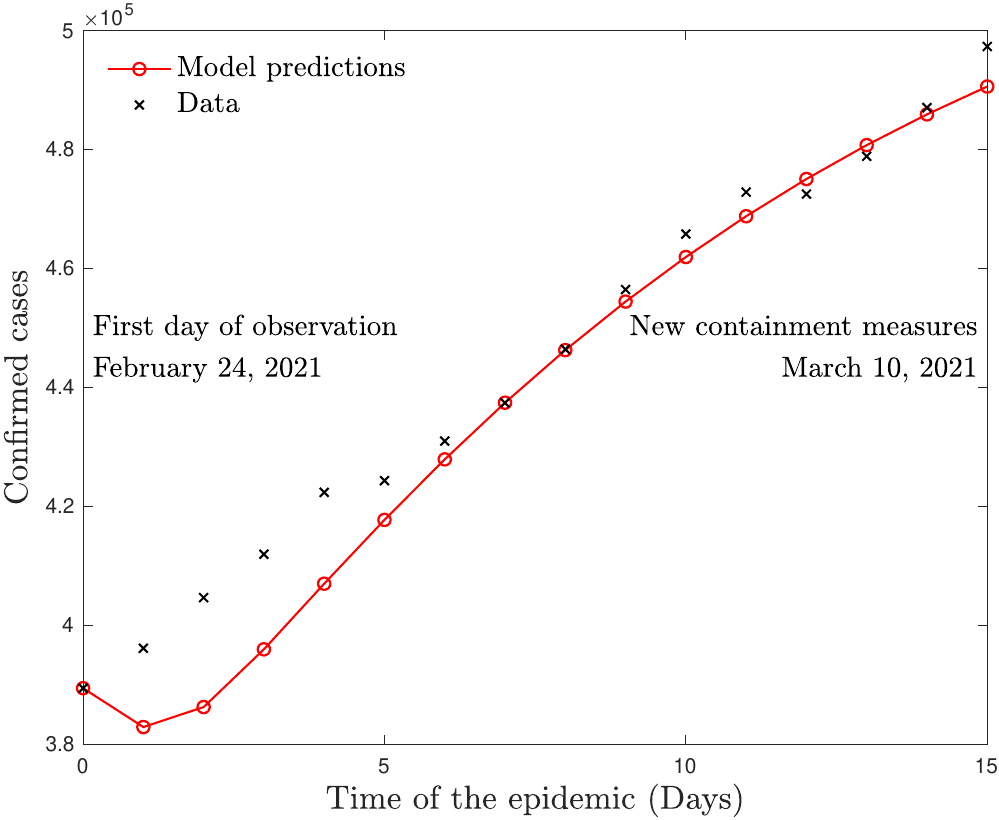}
 \caption{Investigations on the SARS-CoV-2 pandemic in Italy during three distinct epidemic waves, starting from the data collected by Fondazione IRCCS Policlinico San Matteo to infer an initial guess of the infection rate through the function \eqref{eq:VL_shape}. Top row: analysis of the epidemic wave that started in February--March 2020. Optimal shape $\mathbb{E}_z(\beta(\tau,\z))$ of the infectiousness function (left) and model approximation (right) for the cases of infection registered during the period February 24--March 9 2020. Middle row: analysis of the new epidemic outburst that took place in October 2020. Optimal shape of the infection rate (left) and model approximation (right) for the cases of infection registered during the period October 7--22 2020. Bottom row: analysis of the epidemic wave occurred from March 2021.}
 \label{fig:VL Italy}
\end{figure}

The outcomes of our two-step procedure are reported in Figure \ref{fig:VL Pavia} for the case of Pavia and in Figure \ref{fig:VL Italy} for the case of Italy. In the left figures we plot in red the curves providing the mean best shape of the infectiousness function over the three epidemic waves that we have considered. The shaded regions show the $80 \%$ and $95 \%$ percentiles of the different shapes, based on the sample $\mathcal{C}_z$. We then perform numerical simulations to assess the viability of model \eqref{SIR-AOI} to predict the evolution of infected individuals during these subsequent phases of the pandemic. We take the infectiousness function to be the computed mean best shapes over the three periods and we choose $\gamma(\tau,\z)$ constant and equal to 1 for simplicity. In particular, the duration of the infection is solely determined by $\bar{\gamma}$ and the contact rate is given by the optimal ones $\bar{\beta}$ determined from Step 1 and reported in Table \ref{tab:Epidemiological params}. We show in Figures \ref{fig:VL Pavia} and \ref{fig:VL Italy}, on the right, the comparisons between the reported number of infected individuals in Pavia and in Italy over the three epidemic waves and the corresponding approximations provided by model \eqref{SIR-AOI}, with the aforementioned choices of the parameters. We notice that with the progression of the pandemic, specifically over the third wave of February--March 2021, the model loses some accuracy since it is less appropriate to describe slower growths of the epidemic, which stem from the numerous containment measures introduced by the Italian government in the course of 2021 to reduce contacts among the population. This suggests in particular that, in order to capture such intricate effects and minimize the error made by the model, one should consider a variable contact rate $\bar{\beta}(t)$ that continuously evolves over the epidemic time $t$.

\begin{figure}[h!]
\includegraphics[scale=0.4]{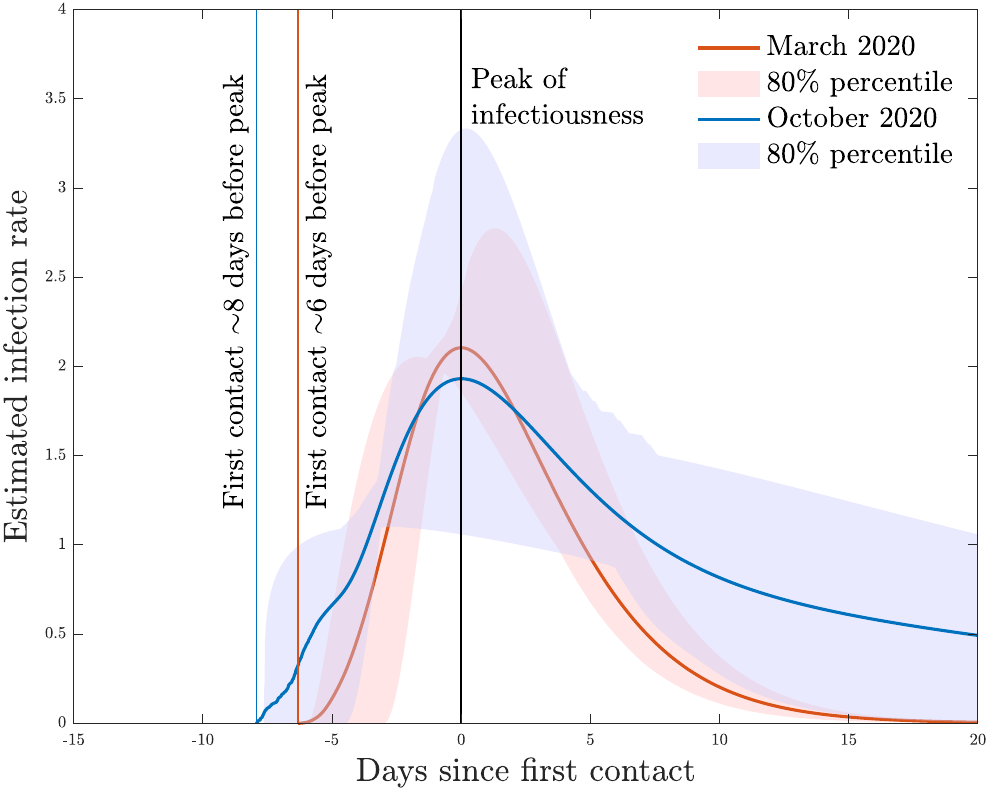} \hspace*{5mm}
\includegraphics[scale=0.4]{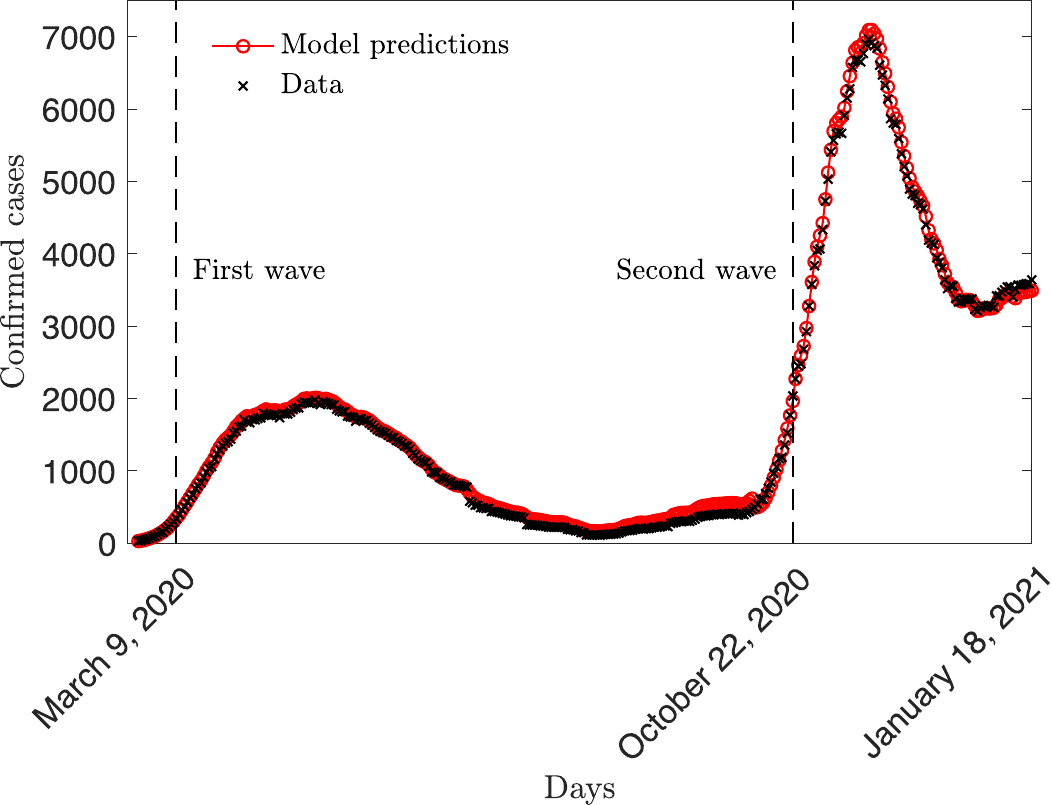}
 \caption{Left: comparison between infection rates from the first (in orange, March 2020) and the second (in blue, October 2020) epidemic waves in the province of Pavia. In order for the model \eqref{SIR-AOI} to best fit the data on confirmed cases, the peak of the infection should be reached about 6 to 8 days after the first contact with an infected subject. Right: evolution of infected individuals from SARS-CoV-2 in Pavia from February 24 2020 to January 18 2021. Comparison between computed (in red) and reported (in black) number of cases.}
 \label{fig:Pavia Whole}
\end{figure}

\begin{figure}[h!]
\centering
\includegraphics[scale=0.4]{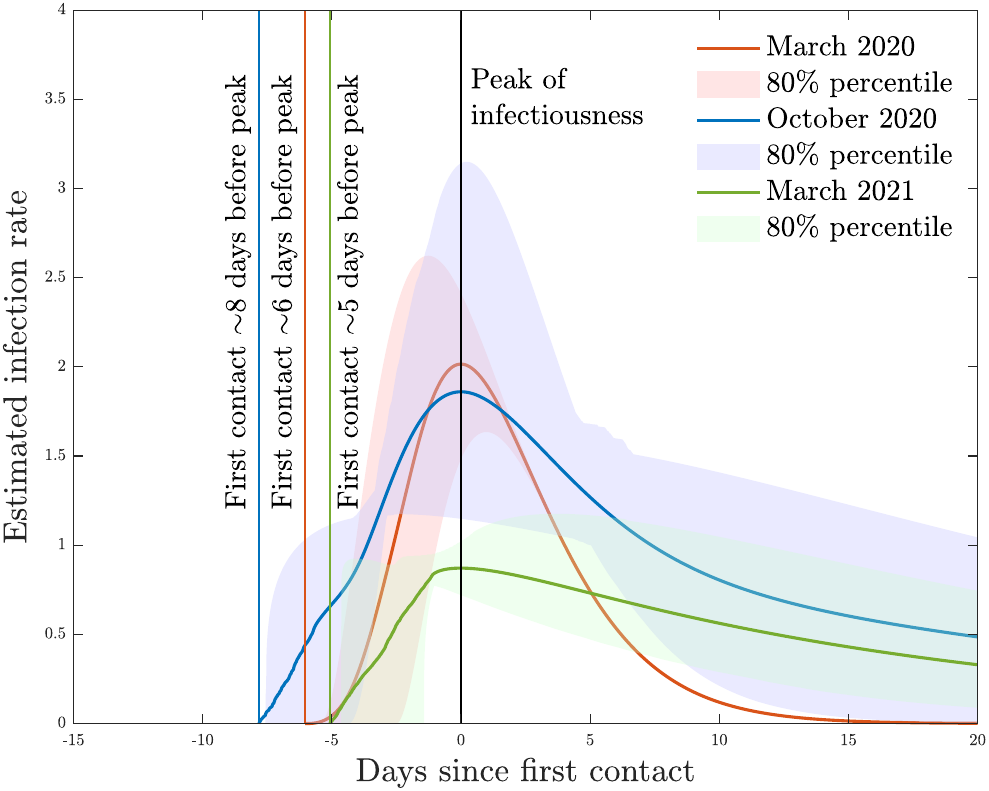} \hspace*{5mm}
\includegraphics[scale=0.4]{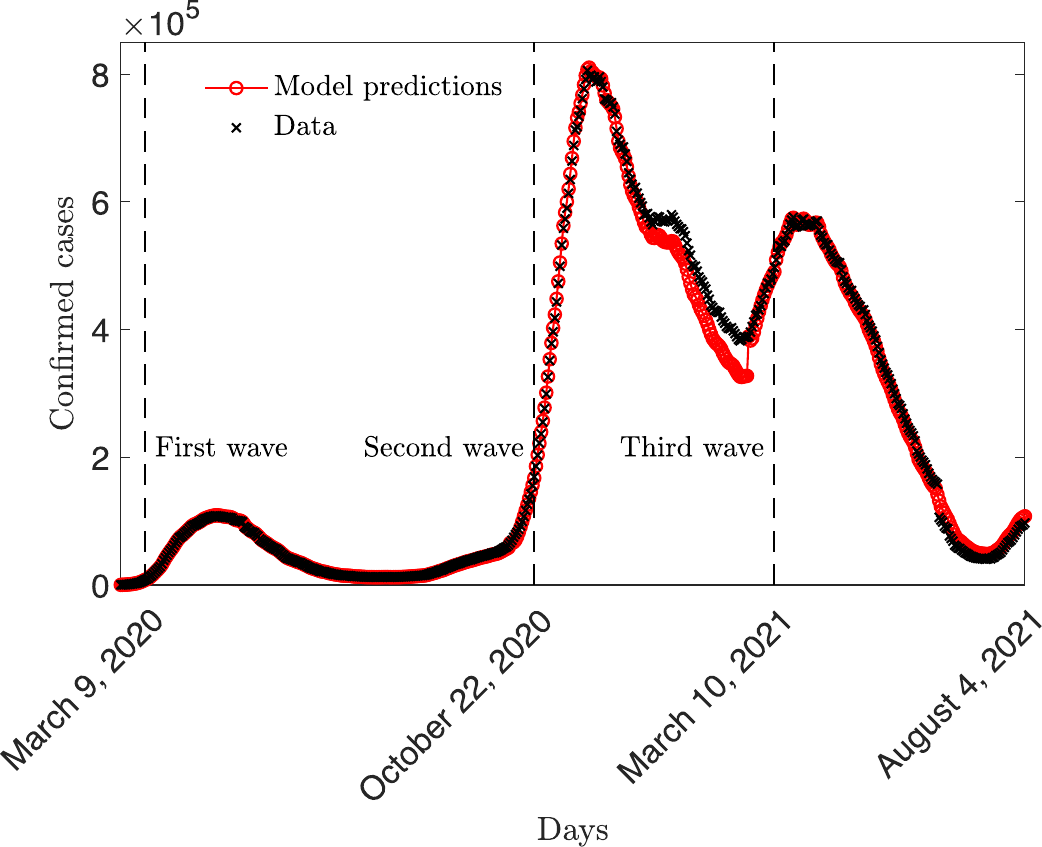}
 \caption{Left: comparison between infection rates from three distinct epidemic waves in Italy happened respectively in March 2020 (orange), in October 2020 (blue) and in March 2021 (green). In order for the model \eqref{SIR-AOI} to best fit the data on confirmed cases, the peak of the infection should be reached about 5 to 8 days after the first contact with an infected subject. Right: evolution of infected from SARS-CoV-2 in Italy from February 24 2020 to August 4 2021. Comparison between computed (in red) and reported (in black) number of cases.}
\label{fig:Italy Whole}
\end{figure}

\smallskip
\noindent \textbf{Reconstructing the experimental data over the whole timeframe.} We now perform a calibration of our model to cover the whole timespan of observation, connecting the epidemic waves between them and providing a full reconstruction of the experimental data on the evolution of infected individuals. We initially observe that, as the epidemiological parameter $\bar{\beta}$ models the contact rate between individuals, its value is highly influenced by the restrictions imposed by a government. This fact needs to be taken into account when extending our model's predictions from one epidemic wave to the other, since various containment measures (we mention in particular the lockdown imposed between March and May 2020) have been put in place by the Italian government over time to slow down the epidemic spreading. Since we do not need to evaluate the shape of the infection rate in these connecting timeframes (the previous two-step analysis is relevant only when the virus is able to spread free among the population, allowing for a proper calibration of the epidemiological parameters), in order to connect the observations we can rely on a simpler model that allows to speed up the computational time. Specifically, we make use of system \eqref{SIR-AOI} with the choices $\beta(\tau,\z), \gamma(\tau,\z) \equiv 1$. From the above considerations, we account for the variable contact restrictions by assuming that the contact rate varies over the epidemic time as $\bar{\beta} = \bar{\beta}(t)$. Then, we determine the value $\bar{\beta}(t)$ by solving an optimization problem for a sequence of time steps $t_i$ (days) over a moving time window of three days (namely, averaging the fitting over three days), varying in any timespan $t \in [t_0+1, t_f]$ where $t_0$ and $t_f$ correspond respectively to the end of an epidemic wave and the beginning of the next one. This strategy is needed in order to correctly assess the evolution of the contact rate depending on the containment measures that were in place. More precisely, we solve the following constrained least-square problem
\begin{equation} \label{Minimization2}
\min_{\bar{\beta}(t_i)} \ (1-\eta) \frac{\norm{ I(t) - \hat{I}(t) }_{L^2([t_i-1,t_i+2])}}{\norm{ \hat{I}(t) }_{L^2([t_i-1,t_i+2])}} + \eta \frac{\norm{ I(t) + R(t) - \hat{I}(t) - \hat{R}(t) }_{L^2([t_i-1,t_i+2])}}{\norm{ \hat{I}(t) + \hat{R}(t) }_{L^2([t_i-1,t_i+2])}},
\end{equation}
where the recovery rate $\bar{\gamma}$ is still kept fixed to the value $1/14$. We complete our study by using the model to reconstruct the experimental data on the epidemic up to the last observation times, being January 18 2021 for the case of the province of Pavia and August 4 2021 for the case of Italy. By means of the above procedure we connect the observations by calibrating $\bar{\beta}(t)$ with the minimization strategy \eqref{Minimization2}. Using all these optimized parameters we can finally run our model over the whole timeframe of observation. We depict the resulting approximations for the evolution of infected individuals in the province of Pavia and in Italy, in Figure \ref{fig:Pavia Whole} and Figure \ref{fig:Italy Whole} respectively, on the right. The evolution of the confirmed cases is plotted in black while the model approximations are shown in red.

At last, in Figures \ref{fig:Pavia Whole} and \ref{fig:Italy Whole} on the left, we provide a direct comparison between the optimal infection rates from different epidemic waves, obtained via the previous two-step analysis, together with their $80\%$ prediction bounds. These are aligned by their respective peak of infectiousness (which corresponds to day $0$ on the graph). One can observe that the infection's peak progressively decreases, while the tails of the infection rate tend to grow. This suggests that the virus becomes weaker but more effective over a longer period of time and individuals stay infective for more days on average. Moreover, the decrease in curve's growth in the beginning of the infection could suggest a slower progression of the symptoms and explain the greater spreading of the epidemic with each new wave.

\section{Conclusion}

Understanding the structure and evolution of VL {dynamics} is crucial in the context of public health when setting up quarantine and isolation measures of positive subjects. In particular, one of the critical issues emerged since the onset of the recent SARS-CoV-2 pandemic has been the need of criteria to assess the infectivity of patients and decide their releasing from prolonged quarantine, safely admitting them back to work duties and social activities. It was demonstrated that high $\textrm{Ct}$ values (low VL), greater than 30-35 $\textrm{Ct}$, were associated with a reduced infectivity of the patients \cite{La Scola,Sin,Pir}. Nevertheless, existing studies indicate that the peak viral load is closely linked with the onset of symptoms, making the available datasets partial and of difficult interpretation. In this work, we have established a pipeline to assess the evolution of in-host VL from respiratory samples of patients. To this end, we combined the use of real data and model-based methods. Furthermore, the proposed approach is capable of tracing the whole VL {dynamics}. In particular, our model helps to provide a data-oriented understanding of the non-infectious status of people that were under quarantine with high $\textrm{Ct}$.

Samples included in this study were collected over two periods of time (November--December 2020 and January--May 2021) that include the initial circulation of VOCs and before the extensive introduction of vaccination. We have first determined the shape of the average VL {dynamics} hinted by these data (recovering Gamma-type distributions)
and we have successively introduced a suitable compartmental model of SIR-type, using a local incidence rate function that varies with the age of infection, to account for the evolution of the epidemic alongside that of the virus. The system of equations belongs to a large class of epidemiological models detailing the transmission dynamics through the inclusion of additional factors such as external influence, age structure, variable contact dynamics and mobility of agents \cite{APZ,BBPSV,CS,DT,KMPH}. In this work, we exploited the information coming from real data on molecular tests and on the reported number of infected individuals to deduce the optimal shape of the infectivity function, based on a combined data-driven and model-driven approach. To take into account the extreme variability of VL dynamics, depending on disease severity and patients' features as well \cite{L_etal,Z_etal}, we also made use of UQ methods \cite{JP,Par,Xiu} adapted for the study of epidemic systems \cite{APZ,CCVH,Chowell,R}, to take into account the uncertainties inherent to the data-assimilation processes of VL kinetics. The model allowed us to infer a plausible evolution of the SARS-CoV-2 infectiousness over three distinct periods of the pandemic (February--March 2020, October 2020 and February--March 2021), in agreement with existing works from the medical literature \cite{CevKupKinPei,PME}. In particular, we observed that the peak of the infection function decreases over time while its tails tend to grow, suggesting that the virus becomes progressively weaker but more effective, since individuals remain infective for more days on average. This change in shape strongly manifests during the third observation period (February--March 2021) when the Alpha variant was actively circulating in Italy, providing an explanation for the greater spreading of the contagion at the time.

In conclusion, our work aimed to implement strategies of preparedness for a new pandemic, since the model and methods developed here could be easily adapted to investigate real data coming from epidemics that are associated with other respiratory viruses (e.g influenza and respiratory syncytial virus), in order to assess the evolution of the associated viral infectiousness that may lead to more informed decisions on preventive measures to reduce the spreading of the disease.


\section*{Acknowledgements}
A.B. and M.Z. acknowledge the support from GNFM of INdAM (National Institute of High Mathematics) and from the Italian Ministry of University and Research (MUR) through the PRIN 2020 project No. 2020JLWP23 (Integrated Mathematical Approaches to Socio-Epidemiological Dynamics). A.B. also acknowledges the support from the European Union's Horizon Europe research and innovation programme, under the Marie Skłodowska-Curie grant agreement No. 101110920 (MesoCroMo - A Mesoscopic approach to Cross-diffusion Modelling in population dynamics). M.Z. wishes to acknowledge partial support by ICSC - Centro Nazionale di Ricerca in High Performance Computing, Big Data and Quantum Computing, funded by European Union - NextGenerationEU.

\begin{figure}[h!]
\includegraphics[scale=0.3]{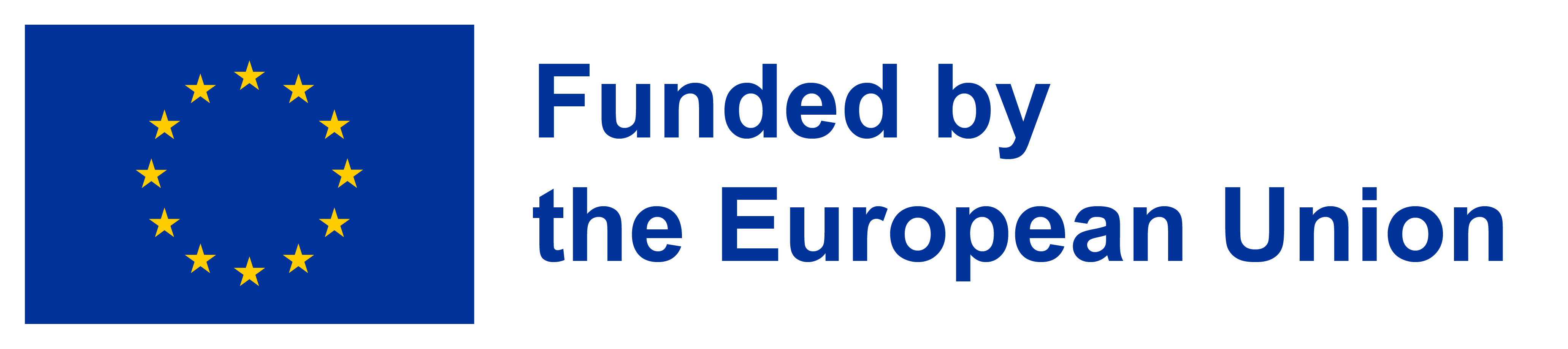}
\end{figure}

\section*{Disclaimer}
Funded by the European Union. Views and opinions expressed are however those of the author(s) only and do not necessarily reflect those of the European Union or of the European Research Executive Agency (REA). Neither the European Union nor the granting authority can be held responsible for them.




\begin{thebibliography}{99}

\bibitem{APZ}
G. Albi, L. Pareschi, M. Zanella. Control with uncertain data of socially structured compartmental epidemic models. \emph{J. Math. Biol.}, \textbf{82}:63, 2021. 

\bibitem{BBPSV}
B. Bartélemy, A. Barrat, R. Pastor-Satorras, A. Vespignani. Dynamical patterns of epidemic outbreaks in complex heterogeneous networks. \emph{J. Theor. Biol.}, \textbf{235}:275--288, 2005.

\bibitem{BBBEPT}
G. Bertaglia, A. Bondesan, D. Burini, R. Eftimie, L. Pareschi, G. Toscani. New trends on the systems approach to modeling {SARS}-{C}o{V}-2 pandemics in a globally connected planet. \emph{Math. Models Methods Appl. Sci.}, \textbf{34}(11):1995--2054, 2024.

\bibitem{BBDP}
G. Bertaglia, W. Boscheri, G. Dimarco, L. Pareschi. Spatial spread of COVI-19 outbreak in Italy using multiscale kinetic transport equations with uncertainty. \emph{Math. Biosci. Eng.}, \textbf{18}(5):7028--7059, 2021.

\bibitem{BPT}
G. Bertaglia, L. Pareschi, G. Toscani. Modelling contagious viral dynamics: a kinetic approach based on mutual utility. \emph{Math. Biosci. Eng.}, \textbf{21}(3): 4241--4268, 2024. 

\bibitem{BBSG}
L. Bolzoni, E. Bonacini, C. Soresina, M. Groppi. Time-optimal control strategies in SIR epidemic models. \emph{Math. Biosci.}, 292:86--96, 2017. 

\bibitem{BTZ}
A. Bondesan, G. Toscani, M. Zanella. Kinetic compartmental models driven by opinion dynamics: Vaccine hesitancy and social influence. \emph{Math. Models Methods Appl. Sci.}, \textbf{34}(6):1043--1076, 2024

\bibitem{Brauer}
F. Brauer, C. Castillo-Chavez, Z. Feng. \emph{Mathematical Models in Epidemiology}, vol. 69, Texts in Applied Mathematics, Springer, New York, 2019. 

\bibitem{BBT}
T. Britton, F. Ball, P. Trapman. A mathematical model reveals the influence of population heterogeneity on herd immunity to SARS-CoV-2. \emph{Science}, \textbf{369}(6505):846--849, 2020. 

\bibitem{BDM}
B. Buonomo, R. Della Marca. Effects of information-induced behavioural changes during the COVID-19 lockdowns: the case of Italy. \emph{R. Soc. Open Sci.}, \textbf{7}(10):201635, 2020. 

\bibitem{CS}
V. Capasso, G. Serio. A generalization of the Kermack--McKendrick deterministic epidemic model. \emph{Math. Biosci.}, \textbf{42}:43--61, 1978.

\bibitem{CCVH}
M. A. Capistr\'an, J. A. Christen, J. X. Velasco-Hern\'andez. Towards uncertainty quantification and inference in the stochastic SIR epidemic model. \emph{Math. Biosci.}, \textbf{240}(2):250--259, 2012. 

\bibitem{CevKupKinPei}
M. Cevik, K. Kuppalli, J. Kindrachuk, M. Peiris. Virology, transmission, and pathogenesis of SARS-CoV-2. \emph{BMJ}, \textbf{371}:1--6, 2020. 


\bibitem{Chen}
J. Chen, H. Lu, G. Melino, et al. COVID-19 infection: the China and Italy perspectives. \emph{Cell Death Dis.}, \textbf{11}:438, 2020. 

\bibitem{Chowell}
G. Chowell. Fitting dynamic models to epidemic outbreaks with quantified uncertainty: a primer for parameter uncertainty, identifiability, and forecasts. \emph{Infect. Dis. Model.}, \textbf{2}(3):379--398, 2017. 

\bibitem{DMLT}
R. Della Marca, N. Loy, A. Tosin. An SIR model with viral load-dependent transmission. \emph{J. Math. Biol.}, \textbf{86}(4):61, 2023. 

\bibitem{DGMM}
J. Demongeot, Q. Griette, Y. Maday, P. Magal. A Kermack--McKendrick model with age of infection starting from a single or multiple cohorts of infected patients. \emph{Proc. R. Soc. A}, \textbf{479}:20220381, 2023. 

\bibitem{DH}
O. Diekmann, J. A. P. Heesterbeek. \emph{Mathematical Epidemiology of Infectious Diseases: Model Building, Analysis and Interpretation}, Wiley, 2000. 

\bibitem{DPTZ}
G. Dimarco, B. Perthame, G. Toscani, M. Zanella. Kinetic models for epidemic dynamics with social heterogeneity. \emph{J. Math. Biol.}, \textbf{83}(1):4, 2021. 

\bibitem{DT}
J. Dolbeault, G. Turinici. Heterogeneous social interactions and the COVID-19 lockdown outcome in a multi-group SEIR model. \emph{Math. Model. Net. Pheno.}, \textbf{15}(36):1--18, 2020. 

\bibitem{FMZ}
J. Franceschi, A. Medaglia, M. Zanella. On the optimal control of kinetic epidemic models with uncertain social features. \emph{Optim. Contr. Appl. Meth.}, \textbf{45}(2):494--522, 2024.

\bibitem{Frediani_etal}
J. K. Frediani, R. Parsons, K. B. McLendon, A. L. Westbrook, W. Lam, G. Martin, N. R. Pollock. The new normal: delayed peak SARS-CoV-2 viral loads relative to symptom onset and implications for COVID-19 testing programs. \emph{Clin. Infect. Dis.}, \textbf{78}(2):301--307, 2024. 

\bibitem{Gatto_etal}
M. Gatto, E. Bertuzzo, L. Mari, S. Miccoli, L. Carraro, R. Casagrandi, A. Rinaldo. Spread and dynamics of the COVID-19 epidemic in Italy: effect of emergency containment measures. \emph{PNAS}, \textbf{117}(19):10484--10491, 2020. 

\bibitem{Giordano}
G. Giordano, F. Blanchini, R. Bruno, P. Colaneri, A. Di Filippo, A. Di Matteo, M. Colaneri. Modelling the COVID-19 epidemic and implementation of population-wide interventions in Italy. \emph{Nat. Med.}, \textbf{26}:855--860, 2020.

{
\bibitem{FIV}
N. Guglielmi, E. Iacomini, A. Viguerie. Identification of time delays in COVID-19 data. \emph{Epidemiol. Meth.}, \textbf{12}(1):20220117, 2023.
}

\bibitem{H_etal}
X. He, E.~H.~Y. Lau, P. Wu, et al. Temporal dynamics in viral shedding and transmissibility of COVID-19. \emph{Nat. Med.}, \textbf{26}:672--675, 2020.

\bibitem{JP}
S. Jin, L. Pareschi. \emph{Uncertainty Quantification for Hyperbolic and Kinetic Equations}, SEMA-SIMAI Springer Series, 14.

\bibitem{J_etal}
T.~C. Jones, G. Biele, B. Mühlemann, et al. Estimating infectiousness throughout SARS-CoV-2 infection course. \emph{Science}, \textbf{373}:eabi5273, 2021.

\bibitem{KDTHM}
M.~J. Keeling, L. Dyson, M.~J. Tildesley, E.~M. Hill, S. Moore. Comparison of the 2021 COVID-19 roadmap projections against public health data in England. \emph{Nat. Commun.}, \textbf{13}(4924), 2022.

\bibitem{KMPH}
M.~J. Keeling, S. Moore, B.~S. Penman, E.~M. Hill. The impacts of SARS-CoV-2 vaccine dose separation and targeting on the COVID-19 epidemic in England. \emph{Nat. Commun.}, \textbf{14}(740), 2023.

\bibitem{K_etal}
E. Kharazmi, M. Cai, X. Zheng, Z. Zhang, G. Lin, G. E. Karniadakis. Identifiability and predictability of integer- and fractional-order epidemiological models using physics-informed neural networks. \emph{Nature Comput. Sci.}, \textbf{1}:744--753, 2021. 

\bibitem{La Scola}
B. La Scola, M. Le Bideau, J. Andreani, et al. Viral RNA load as determined by cell culture as a management tool for discharge of SARS-CoV-2 patients from infectious disease wards. \emph{Eur. J. Clin. Microbiol. Infect. Dis.}, \textbf{39}(6):1059--1061, 2020.

\bibitem{Lavezzo}
E. Lavezzo, E. Franchin, C. Ciavarella, et al. Suppression of a SARS-CoV-2 outbreak in the Italian municipality of Vo'. \emph{Nature}, \textbf{584}:425--429, 2020.

\bibitem{L_etal}
Y. Liu, L.-M. Yan, L. Wan, et al. Viral dynamics in mild and severe cases of COVID-19. \emph{Lancet Infect. Dis.}, \textbf{20}(6):656--657, 2020.

\bibitem{LT}
N. Loy, A. Tosin. A viral load-based model for epidemic spread on spatial networks. \emph{Math. Biosci. Eng.}, \textbf{18}(5):5635--5663, 2021.

\bibitem{LGT}
H. Lu, F. Giannino, D.~M. Tartakovsky. Parsimonious models of in-host viral dynamics and immune response. \emph{Appl. Math. Lett.}, \textbf{145}:108781, 2023.

\bibitem{Medaglia_etal}
A. Medaglia, G. Colelli, L. Farina, A. Bacila, P. Bini, E. Marchioni, S. Figini, A. Pichiecchio, M. Zanella. Uncertainty quantification and control of kinetic models of tumour growth under clinical uncertainties. \emph{Int. J. NonLin. Mech.}, \textbf{141}:103933, 2022.


\bibitem{M_etal}
M. Mozgovoj, M.~D. Graham, C. Ferrufino, S. Blanc, A.~F. Souto, M. Pilloff, M.~J. Dus Santos. Viral load in symptomatic and asymptomatic patients infected with SARS-CoV-2. What have we learned? \emph{J. Clin. Virol. Plus}, \textbf{3}(4):100166, 2023. 


\bibitem{Par}
L. Pareschi. An introduction to uncertainty quantification for kinetic equations and related problems. In G. Albi, S. Merino-Aceituno, A. Nota, M. Zanella(eds.) \emph{Trails in Kinetic Theory: Foundational Aspects and Numerical Methods}, SEMA-SIMAI Springer Series, vol. 25.


\bibitem{PD}
A. Piralla, C. Daleno, E. Pariani, P. Conaldi, S. Esposito, A. Zanetti, F. Baldanti. Virtual quantification of influenza A virus load by real-time RT-PCR. \emph{J. Clin. Virol.}, \textbf{56}(1):65--68, 2013. 

\bibitem{Pir2}
A. Piralla, A. Girello, M. Premoli, F. Baldanti. A New Real-Time Reverse Transcription-PCR Assay for Detection of Human Enterovirus 68 in Respiratory Samples. \emph{J. Clin. Microbiol.}, \textbf{53}(5): 1725--1726, 2015.


\bibitem{Pir}
A. Piralla, M. Ricchi, M.~G. Cusi, et al. Residual SARS-CoV-2 RNA in nasal swabs of convalescent COVID-19 patients: Is prolonged quarantine always justified? \emph{IJID}, \textbf{102}:299--302, 2021.

\bibitem{P_etal}
O. Puhach, K. Adea, N. Hulo, et al. Infectious viral load in unvaccinated and vaccinated individuals infected with ancestral, Delta or Omicron SARS-CoV-2. \emph{Nat. Med.}, \textbf{28}:1491--1500, 2022. 

\bibitem{PME}
O. Puhach, B. Meyer, I. Eckerle. SARS-CoV-2 viral load and shedding kinetics. \emph{Nat. Rev. Microbiol.}, \textbf{21}:147--161, 2023. 

\bibitem{R}
M. G. Roberts. Epidemic models with uncertainty in the reproduction number. \emph{J. Math. Biol.}, \textbf{66}:1463--1474, 2013.

\bibitem{Sin}
A. Singanayagam, M. Patel, A. Charlett, et al. Duration of infectiousness and correlation with RT-PCR cycle threshold values in cases of COVID-19, England, January to May 2020. \emph{Euro Surveill.}, \textbf{25}(32):2001483, 2020.

\bibitem{V}
R. Verity, L.~C. Okell, I. Dorigatti, et al. Estimates of the severity of coronavirus disease 2019: a model-based analysis. \emph{Lancet Infect. Dis.}, \textbf{20}(6):669--677, 2020.

\bibitem{Xiu}
D. Xiu. \emph{Numerical Methods for Stochastic Computations: a Spectral Methods Approach}. Princeton University Press, Princeton, 2010.

\bibitem{YZW}
J.-Y. Yang, F.-Q. Zhang, X.-Y. Wang. SIV epidemic models with age of infection. \emph{Int. J. Biomath.}, \textbf{2}(1):61--67, 2009.

\bibitem{Z_BMB}
M. Zanella. Kinetic models for epidemic dynamics in the presence of opinion polarization. \emph{Bull. Math. Biol.}, \textbf{85}:36, 2023.

\bibitem{Z_etal}
M. Zanella, C. Bardelli, G. Dimarco, S. Deandrea, P. Perotti, M. Azzi, S. Figini, G. Toscani. A data-driven epidemic model with social structure for understanding the COVID-19 infection on a heavily affected Italian Province. \emph{Math. Models Methods Appl. Sci.}, \textbf{31}(12):2533--2570, 2021.


\end{thebibliography}
\end{document}